# Superconductivity in ultra-thin carbon nanotubes and carbyne-nanotube composites: an ab-initio approach

[1]C. H. Wong, E. A. Buntov, R. E. Kasimova, A. F. Zatsepin

Institute of Physics and Technology, Ural Federal University, Yekaterinburg, Russia

**Abstract:**

The superconductivity of the 4-angstrom single-walled carbon nanotubes (SWCNTs) was discovered more than a decade ago, and marked the breakthrough of finding superconductivity in pure elemental undoped carbon compounds. The van Hove singularities in the electronic density of states at the Fermi level in combination with a large Debye temperature of the SWCNTs are expected to cause an impressively large superconducting gap. We have developed an innovative computational algorithm specially tailored for the investigation of superconductivity in ultrathin SWCNTs. We predict the superconducting transition temperature of various thin carbon nanotubes resulting from electron-phonon coupling by an ab-initio method, taking into account the effect of radial pressure, symmetry, chirality (N,M) and bond lengths. By optimizing the geometry of the carbon nanotubes, a maximum $T_c$ of 60K is found. We also use our method to calculate the $T_c$ of a linear carbon chain embedded in the center of (5,0) SWCNTs. The strong curvature in the (5,0) carbon nanotubes in the presence of the inner carbon chain provides an alternative path to increase the $T_c$ of this carbon composite by a factor of 2.2 with respect to the empty (5,0) SWCNTs.

**Introduction:**

The observation of superconductivity in 4-angstrom single wall carbon nanotubes arrays (SWCNTs) was first reported back in 2001 [1], and further confirmed by more detailed experiments [2]. The superconducting transition temperature $T_c$ was observed with onset at 15K. In a purely one-dimensional (1D) material the electrons form Cooper pairs at the onset superconducting transition temperature, however the low-dimensionality causes strong thermal and quantum fluctuations leading to phase slips events, which cause finite resistance at any non-zero temperature [3]. This scenario has been theoretically well described in the framework of the Langer-Ambegaokar-McCumber-Halperin (LAMH) theory [3,4]. Fortunately, this limitation of 1D superconductors can be overcome by arranging the superconducting nano-elements in the form of closely packed arrays of parallel wires [1,5,6,7,10]. The Josephson interaction induced by quantum tunneling of Cooper pairs stabilizes then the superconducting phase order parameter and triggers a dimensional crossover from a 1D fluctuating state at high temperatures to a 3D phase coherent state with vanishing electrical resistance [5,6,7,8,9,10]. By drawing a parallel to the geometry of the non-superconducting graphene, a strong curvature of the graphene sheet is believed to be the main ingredient to activate the superconductivity in thin carbon nanotubes [1,11]. The BCS theory states that a large electronic density of states (DOS) at the Fermi level and a high Debye frequency are the essential ingredients for a high $T_c$ in the case of classical phonon-mediated BCS superconductors [8]. 1D metallic elements feature van Hove singularities in their electronic density of states (DOS) at Fermi level, and if by chance such a singularity could appear at, or in the vicinity of the Fermi level very high $T_c$ values could result. The Fermi level may be further tuned by application of pressure or electric gate voltages [12], and in addition the superconductivity of SWCNTs may be further tunable through the lateral tube-to-tube distances and bond lengths. However, the synthesis of high quality carbon nanotubes thinner than 4 Angstrom in diameter represents a challenge [13]. In addition, no theoretical model predicting the superconducting transition temperature $T_c$ of the SWCNTs accurately was reported so far, despite of numerous first principle calculations that have been reported for SWCNTs [14,15]. In view of this, we have developed a powerful theoretical model to accurately predict superconducting parameters of thin carbon nanotubes. In addition, we address linear carbon chains, whose existence has been questioned for a long time due to its energetic instability [16]. However, linear carbon nanowires protected by double walled carbon nanotube (DWCNT) have been fabricated successfully recently [16], and we will study how the interaction between the chain and the nanotube can cause superconductivity in such a carbon composite. In our simulation the carbon nanowire is surrounded by a (5,0) SWCNTs. The (5,0) SWCNTs features a threshold radius, which means that no extra covalent bond is established radially between the carbon nanowire and the nanotube.

[1] ch.kh.vong@urfu.ru

**Computational Methods**

The BCS pairing Hamiltonian, $H_{pair} = \sum_{k\sigma} E_k n_{k\sigma} + \sum_{kl} V_{kl} c^*_{k\uparrow} c^*_{k\downarrow} c_{l\uparrow} c_{l\downarrow}$, is made up of the single particle energy $E_k$ relative to the Fermi energy. The interaction $V_{kl}$ changes the state of particle from $(l\uparrow,-l\downarrow)$ to $(k\uparrow,-k\downarrow)$. The creation operators, $c^*_{k\uparrow}$ and $c^*_{k\downarrow}$, refer to spin up and down respectively, while the particle number operator is represented by $n_{k\sigma}$ and $\sigma$ is the spin index [8]. The ground state of the BCS wavefunction $\psi_G$ is expressed as

$$|\psi_G\rangle = \prod_{k=k_1...k_M} \left(u_k + v_k c^*_{k\uparrow} c^*_{k\downarrow}\right)|\varphi_0\rangle$$

where $|\varphi_0\rangle$ is the vacuum state with the absence of particles. As $|u_k|^2 + |v_k|^2 = 1$, the $|u_k|^2$ means the unoccupied probability. In BCS material the energy gap is $k$ independent and hence we may define $\Delta = \Delta_k = -\sum_l V_{kl} u_l v_l$ [8]. The interaction term is originated from the electron phonon scattering in the expression of $H_{e-ph} = \sum_{kk'\sigma\lambda} g_{kk'\lambda} C^\dagger_{k\sigma} C_{k'\sigma}((a_\lambda(\mathbf{q}) + a^\dagger_\lambda(-\mathbf{q}))$ [8,11]. The $C_{k\sigma}$ and $\lambda$ result from linear combinations of the eigenfunctions and polarizations. The $g_{kk'\lambda}$ is related to the electrostatic integral and lattice vibrations. The $a_{q\lambda}^{\dagger} a_{q\lambda} + \frac{1}{2}$ refers to the quantum number of the phonons, where $\mathbf{q} = \mathbf{k} - \mathbf{k}' + \mathbf{G}$ and $\mathbf{G}$ correspond to a reciprocal lattice vector [12].

For circular materials like SWCNT, the attractive force acting on the electrons needs to be modified due to the increase of the effective atomic number $Z_{effective}$ [17].

$$Z_{effective} = Z \frac{\sum_{r}^{R} U_c(r)}{\sum_{r}^{R} U_p(r)}$$

The Bloch theorem ensures that the wavefunctions of electron $\psi$ can be written in the form of $\psi(r+R) = e^{ik \cdot R} \psi(r)$ where $k$ is the wave number and $R$ is a lattice vector. By comparing the attractive potential between the circular $U_c$ and planar $U_p$ shapes, the $Z_{effective}$ is obtainable.

The prediction of the $T_c$ is acquired by computing the scale factor $\frac{T_{c(A)}}{T_{c(B)}} = \frac{\Delta_A(0)}{\Delta_B(0)}$, because $\Delta(0) \propto T_c$ per electron [8]. If the $T_c$ of the material B is known, the $T_c$ of the material A is predictable according to our semi-phenomenological scale-factor approach. However, the $u_l v_l$ depend on $\Delta$ and hence another transfer function is needed: According to the BCS theory, the energy gap is expressed as $\Delta_k = -\frac{1}{2} \sum V_{kl} \frac{\Delta_l}{(\Delta_l^2 + E_k^2)^{0.5}}$ [8]. The derivation of the transfer function starts from calculating the trial energy gap $\Delta^T$, which originates from the electrons at the Fermi level only. In this particular situation the $\Delta^T$ is directly proportional to the interaction term after the vanishing of $E_k$. Then the transfer function, i.e. $u_l^T v_l^T$ as a function of electron energy is interpreted, which will be substituted into $\Delta_k^{corrected} = -\sum_{kl} V_{kl} u_l^T v_l^T$ to correct the energy gap. However, the scale factor approach is valid if it satisfies the condition that the Debye energy $\hbar \omega_D \gg \Delta$. Otherwise, the BCS occupational fraction will not drop sharply to zero when the electron energy increases [8].

We apply the scale factor approach to calculate the $T_c$ of SWCNTs arranged in the form of a hexagonal array. Each SWCNT can be imagined as a rolled-up graphene sheet forming the shape of a tube. As a result, the phonon wavefunction in a SWCNT $\chi(\omega_x^{planar}, \omega_y^{planar}, \omega_z^{circular})$ becomes related to graphene $\chi(\omega_x^{planar}, \omega_y^{planar})$ and linear carbon nanochains $\chi(\omega_x^{linear})$ of the same bond lengths. We consider two types of repeating units in graphene as illustrated in Figure 1. The effective spring constants $(K_x^{planar}, K_y^{planar})$ of the upper and lower repeating units in graphene are resolved into the $x$ and $y$ axis, respectively.

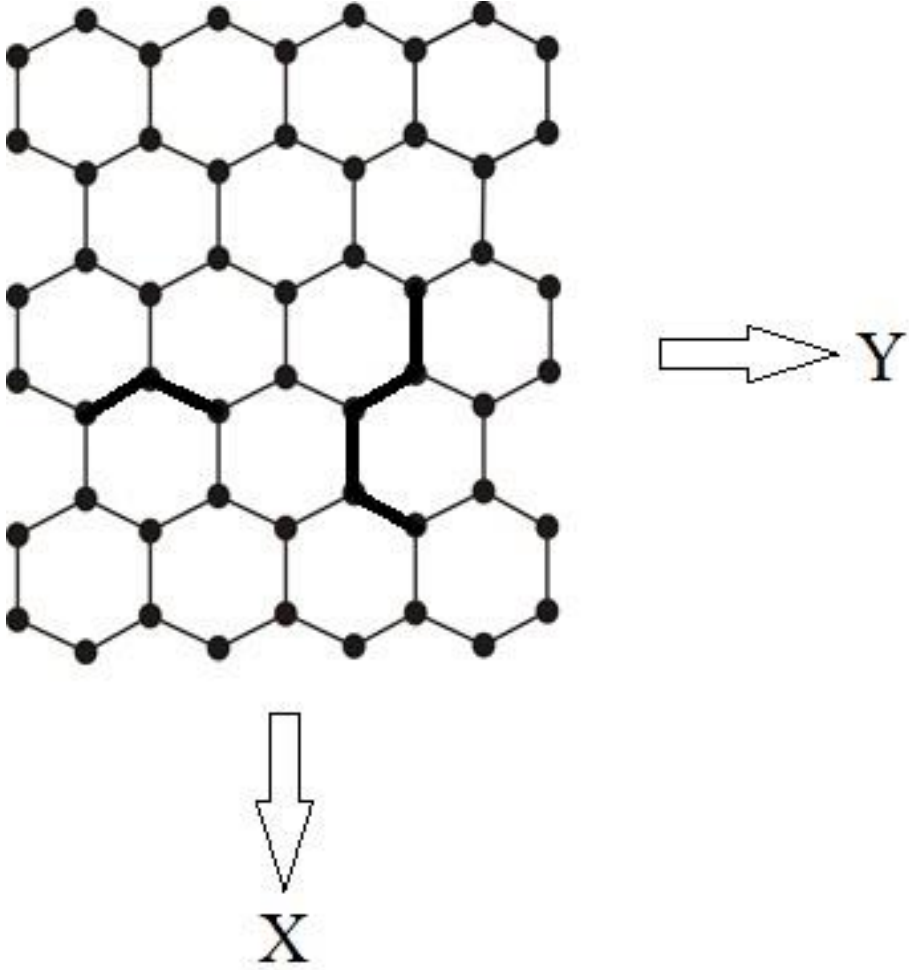

**Figure 1:** Illustration of the structure of graphene. Two types of repeating units of graphene are shown in thicker lines.

As the bond length of graphene is about 143 pm, we make use of the GGA functional [18,19] to simulate the dispersion curve and the phonon density of states in the linear carbon chain under the same bond distance of 143 pm based on the finite displacement method in which the corresponding supercell cut-off radius is 0.5nm [20]. In addition, the electronic DOS of the reference carbon chain is simulated by the GGA functional in the Dmol$^3$ package [21]. The lateral chain-to-chain separation between the isolated carbon chains is 1340 pm. By comparing the linear repeating unit made by the four nearest carbon atoms along the reference linear carbon chain with the known dimensionless spring constant $K_x^{linear}$, the vibrational frequency of the graphene is interpretable after computing the ratio of the resultant spring constant of the graphene relative to the reference chain using the classical mass-spring formula of

$$\frac{\sqrt{(K_x^{planar} K_x^{planar} + K_y^{planar} K_y^{planar})}}{K_x^{linear}}$$

[12]. In other words, the approximated phonon wavefunction in graphene can be estimated by the lowest order of the Hermit polynomial [12]. Using the ground state

Hermit polynomial is sufficiently appropriate due to the low temperatures [12]. In the next step the approximated wavefunction of the SWCNT is calculated with help of the concept of curvature-assisted phonon softening [11]. For instance, a loop in the (4,2) SWCNT is formed by about 13 atoms along the armchair path, and eventually the tilt angle between the adjacent carbon atoms is 360/13 = 27.6 degrees. Another relative spring constant $K_z^{circular}$ of the SWCNT is necessary to obtain the full approximated wavefunction of the SWCNT, which can be found by resolving the lattice vibration into radial and tangential components [11].

The electronic band diagram and the electronic DOS of the SWCNTs are determined by the GGA functional [18] in the CASTEP directly. The energy cut-off point and tolerance are 200eV and 10μeV, respectively. The potential energy in the Schrodinger equation of electrons is found by projecting the electric fields onto the orthogonal directions, and the time-independent Schrödinger equation of electrons is solvable via the method of separation of variables [12]. Finally, the theoretical $T_c$ of the (N,M) SWCNTs can be found. The predicted $T_c$ of the SWCNT will be calculated relative to aluminum, lead, tin, indium, mercury and tantalum respectively [22]. The aim is to ensure all predicted $T_c$ values of the SWCNTs based on the scale factor approach are almost identical, despite of the choice of different reference materials involving various lattice structures. The effective Coulomb coupling strengths of the BCS materials [23] are calculated from the Fermi energy, Debye energy, Thomas Fermi wave number and the Fermi wave number.

The conclusive equation of the BCS theory [8], i.e. $\Delta = \dfrac{\hbar \omega_D}{\sinh\left[1/DOS(E_F)V\right]}$, is obtained by considering the finite integral of the electron energy up to the Debye frequency. If the 1D material contains extremely narrow peaks in the DOS, the difference between the $DOS(E_F)$ and $DOS(E_F + dE)$ may become very large [3], and hence the formula may become inapplicable. Therefore, we do not substitute the physical quantities directly into the conclusive formula, because this may not work in some 1D materials with a high Debye frequency. The accuracy of the transfer function is acceptable for aluminum, lead, tin, indium, mercury, tantalum and SWCNT, because their Debye temperatures are much higher than the observed superconducting transition temperatures [12,22]. The argument is based on the BCS occupational probability, which will drop quickly to zero beyond the Fermi level if the Debye energy $\hbar\omega_D$ is infinitely large [8], which more or less matches the pattern of the transfer function. A careful comparison between the errors of the survival energies of the paired electrons is made. Assuming that the $\Delta^T$ of the (4,2) SWCNTs is three times higher at the same Debye temperature, the average offset between $\left\langle \sum_{E_F}^{E_F+\Delta^T} u_l v_l \right\rangle$ and $\left\langle \sum_{E_F}^{E_F+3\Delta^T} u_l v_l \right\rangle$ is still less than 4% when we sum over the survival energies of the paired electrons in a material with large Debye temperature. The electronic properties of the carbon chain surrounded by (5,0) SWCNTs is calculated by the GGA functional [18] in CASTEP and the corresponding phonon properties are acquired by the finite displacement algorithm again [20]. The data of the carbon composite is implemented in the scale factor approach again.

**Results:**

The predicted $T_c$ of the (4,2) and (5,0) SWCNTs separated by ~0.75nm are 13.7K and 8.1K respectively. Our proposed computational model shows good agreement to the experimental $T_c$ of the mixed (4,2) & (5,0) single walled carbon nanotube arrays [24]. Figure 2 display the energy band gap of the (4,2) SWCNTs as a function of tube-to-tube distance, where the nanotube diameter is 0.39nm. The semiconductor-metal transition takes place at a separation between 0.7 and 0.9nm. Figure 3 shows that the electronic DOS at the Fermi level in the (2,2) SWCNTs increases with the tube-to-tube spacing. However, the increase in the DOS saturates once the nanotubes reach a separation exceeding 0.8nm, where the maximum DOS corresponds to ~0.43 electron/eV per atom. Apart from this, we have verified that the theoretical $T_c$ of the (2,2) SWCNTs also remains unchanged at ~36K in the saturated range. As shown under 'category A' in Table 1, the weakening effect on $T_c$ is observed at narrower tube-to-tube distance where the $T_c$ of the (2,2) SWCNTs separated by 0.6nm is reduced to 22K.

Will asymmetric alignment of the nanotubes array increase the superconducting energy gap? Unfortunately, the theoretical $T_c$ of 'category B' demonstrates that the asymmetric arrangement of the (2,2) SWCNTs provides no extra reinforcement of $T_c$. Table 2 compares the electronic DOS between the elliptical and circular (2,2) SWCNTs subjected to radial pressure, where the lateral tube-to-tube spacing is fixed to 0.305nm and 0.605nm, respectively. The radially stressed (2,2) SWCNTs reduces the electronic DOS at the Fermi level by 57%, significantly. However, applying pressure in the correct way is naively expected to raise the $T_c$, and indeed evidence is provided in Table 3. The $T_c$ of the (2,2) SWCNTs is enhanced from 35.6K to 59.8K after reducing the covalent bond length from 143pm to 134pm.

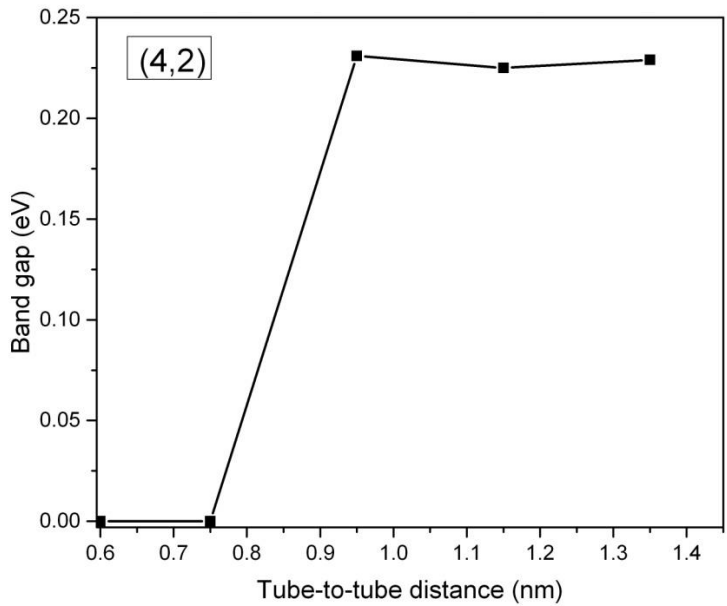

Figure 2: The energy band gap of the (4,2) SWCNTs as a function of tube-to-tube separation.

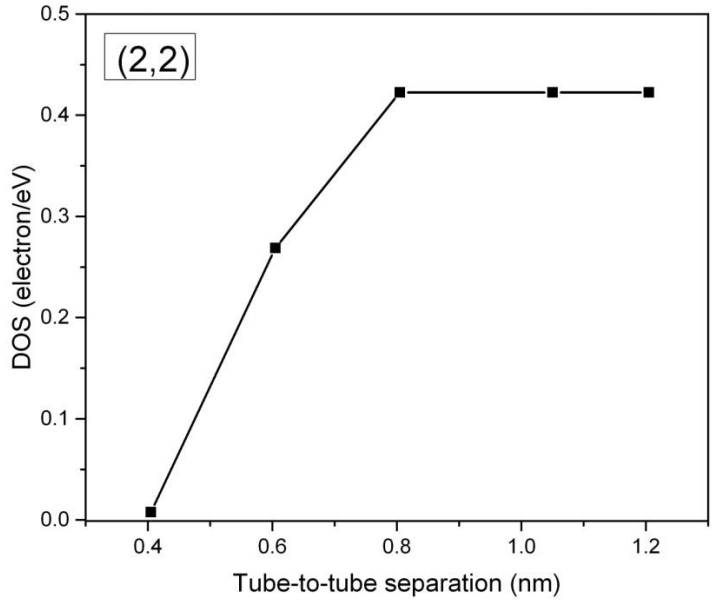

Figure 3: The electronic DOS at Fermi level per atom of the (2,2) SWCNTs as a function of tube-to-tube separation. The diameter of (2,2) SWCNTs is 0.27nm.

Table 1: The effect of asymmetric tube-to-tube separations on the superconductivity of (2,2) SWCNTs. The lateral tube-to-tube distances, $L_x$ and $L_y$, are perpendicular to long axis

| Category | $L_x$/nm | $L_y$/nm | $T_c$/K |
|---|---|---|---|
| A | 0.605 | 0.605 | 21.8 |
| B | 0.605 | 1.05 | 29.6 |
| C | 1.05 | 1.05 | 35.6 |

Table 2: Radial pressure effect on the electronic DOS at the Fermi level for (2,2) SWCNTs

| Major axis/nm | Minor axis/nm | DOS per atom |
|---|---|---|
| 0.137 | 0.137 | 0.202 (circular shape) |
| 0.233 | 0.121 | 0.086 (Elliptical shape) |

Table 3: The influence of bond length on the superconductivity of isolated (2,2) SWCNTs.

| Bond length/pm | $T_c$/K |
|---|---|
| 143 | 35.6 |
| 134 | 59.8 |

Table 4: The $T_c$ of various zig-zag SWCNTs of different diameters. The tube-to-tube separation is about 0.75nm.

| (N,M) | $T_c$ /K |
|---|---|
| (3,0) | 26.5 |
| (4,0) | 37.1 |
| (5,0) | 8.1 |

We further investigate the curvature effect on the superconductivity of zigzag SWCNTs. In comparison to (3,0) and (5,0) SWCNTs, the $T_c$ of (4,0) SWCNTs is the highest in the zigzag family, as demonstrated in Table 4. The aim of Figure 4 is to find out the minimum tube-to-tube separation to isolate the individual SWCNTs, based on the variation of the electronic DOS. The threshold lateral separation of the (3,0) and (4,0) SWCNTs are almost identical at 0.75nm. However, the threshold distance of the (5,0) SWCNTs is slightly larger. The DOS of the (3,0), (4,0), (5,0) SWCNTs remain constant once the nanotubes become separated by more than 1nm. Figure 5 provides the energy scan of the DOS. Through growing the linear carbon chain along the (5,0) SWCNTs, the relaxed structure of the carbon composite is shown in Figure 6. Obviously, the inner carbon nanowire is twisted from linear to a zigzag form in order to minimize its energy. Interestingly, the covalent bond distances of the zigzag carbon nanowire are 1.578Å and 1.404 Å alternatively with the bond angle of 135.634 degree. Figure 7 confirms the metallic state of the carbon composite, which still stands a chance to become superconducting. Simulating the electronic DOS distribution of the carbon composite provides 0.23 electron / eV at the Fermi level in Figure 8. The electronic DOS of the carbon composite at the Fermi level is ~1.5 times larger than the pure (5,0) SWCNTs under the same lateral spacing of 1.7nm. The average phonon frequency of the composite is ~46 rad/s. With these parameters we obtain a predicted $T_c$ of this carbon composite of 26K.

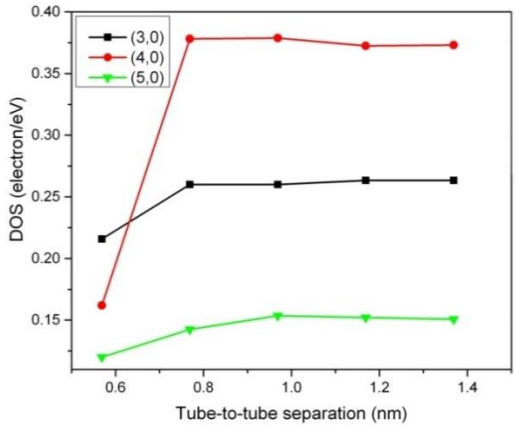

Figure 4: The electronic DOS at the Fermi level per atom of the (3,0), (4,0) and (5,0) SWCNTs as a function of tube-to-tube interaction. The diameter of (3,0), (4,0), (5,0) SWCNT are 0.24nm, 0.31nm and 0.39nm, respectively.

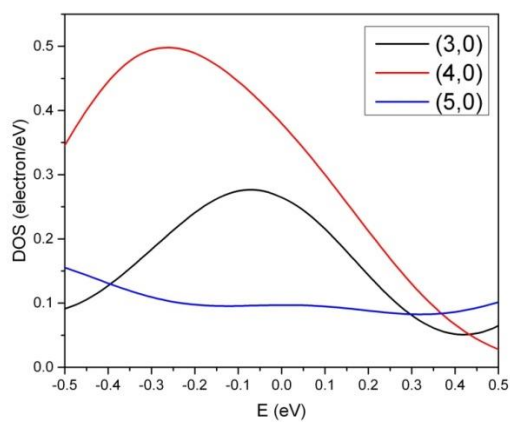

Figure 5: Electronic density of states per atom of zigzag SWCNTs of various radiuses. The Fermi level is shifted to 0eV for clarity. The lateral tube-to-tube distance is about 0.75nm

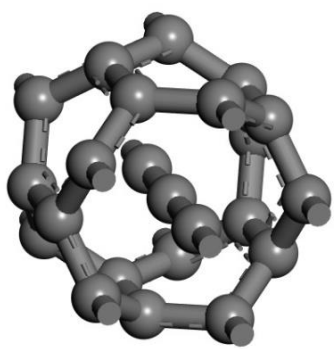
Before geometric optimization

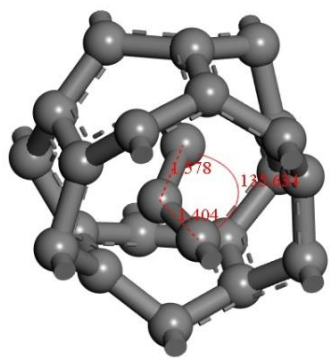
After geometric optimization

Figure 6: The carbon composite is made by inserting a linear carbon chain into a (5,0) SWCNT. The LHS shows the input atomic coordinates of the repeated unit, while the RHS displays the repeated unit after geometric optimization. The composites are arranged to form a regular array in which they are laterally separated by 1.7nm.

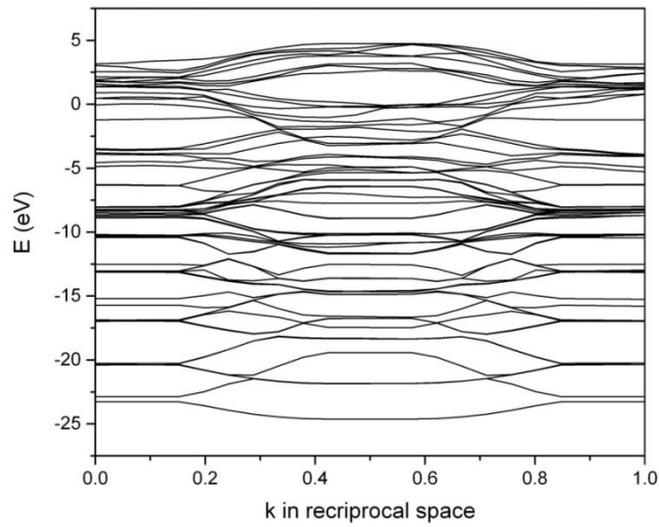

Figure 7: Electronic band diagram of the carbon composite arrays (linear carbon chains surrounded by (5,0) SWCNTs) where the lateral tube-to-tube distance is 1.7nm. The Fermi level is shifted to 0eV for clarity.

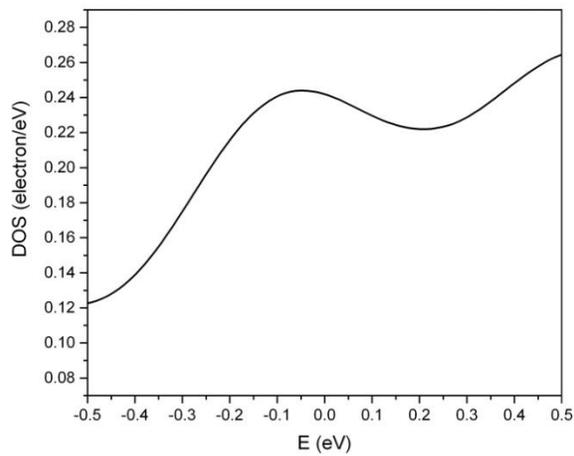

Figure 8: Electronic DOS per atom of the carbon composite (linear carbon chain embedded in (5,0) SWCNTs). The composites are separated by 1.7nm.

**Discussion:**

Although it is widely believed that the ultra-thin (4,2) SWCNTs should be semiconducting, a semiconductor-metal transition of the (4,2) SWCNTs arranged in parallel arrays occurs for tube-to-tube separation of less than ~0.9nm, which is due to the more effective orbital overlap. In experimental work on superconducting SWCNT arrays grown in $AlPO_4$-5 zeolite matrices containing mixed (4,2) and (5,0) SWCNTs [1], the experimental onset $T_c$ at 15K is likely attributed to the (4,2) SWCNTs, in contrast to what has been proposed previously [25], and the sharper resistance downturn at ~7K in zero magnetic field [1] is likely credited from the onset $T_c$ in the (5,0) SWCNTs. The superconducting (4,2) and (5,0) nanotubes simultaneously reinforce the transverse Josephson interaction within the array to trigger a dimensional transition towards a three-dimensional bulk superconducting state within the array [5,6,9]. The isolated (2,2) SWCNTs carries a larger electronic DOS, which is due to the presence of the Van Hove singularity in the 1D regime, as shown in Figure 3. The change of the attractive potential here influences the electron wavefunction that causes the variation in the DOS [3,12]. A poor sample quality in terms of irregular alignment of SWCNTs will lower the $T_c$ as shown in Table 1.

Generating elliptical SWCNT can be achieved by applying pressure radially. However, compressing the carbon nanotube array not only change the shape, but also change the tube-to-tube separation asymmetrically, and therefore we set two different tube-to-tube spacings, 0.305nm and 0.605nm, respectively, to compare the electronic DOS between the circular and elliptical (2,2) SWCNTs in Table 2. The reduction of the electronic DOS in the elliptical shape can be explained by the atomic density. If the semi-minor axis is much smaller than the semi-major axis, the carbon atoms are occupied denser [26] and then the electronic DOS is lowered with the proof provided in Figure 3. In view of this, the increase of $T_c$ due to the elliptical shape is likely out of our expectation. However, the $T_c$ of the (2,2) SWCNT is increased up to 60K by controlling the bond length to 134pm as shown in Table 3. It is mainly credited from the stronger Bloch potential resulting from the shorter bond distance and the simultaneous enhancement of the electronic DOS at the Fermi level [8].

Thinner carbon nanotube are generally expected to have a higher $T_c$, but Table 4 provides evidence that this belief is macroscopically true only, where the optimal $T_c$ is observed in (4,0) SWCNTs in the zigzag family. Although the phonon softening due to the curvature in the (4,0) SWCNTs is weaker than for the (3,0) SWCNTs, the increase of the electronic DOS in the (4,0) nanotubes outweighs the effect of phonon softening as demonstrated in Figure 4. Another interesting phenomenon is that the (5,0) nanotubes hold a longer cut-off distance for the lateral interaction to vanish as shown in Figure 5. To investigate this phenomenon, it is necessary to convert the tube-to-tube distance to a wall-to-wall distance. Assuming that the tube-to-tube distance is 1nm, the (3,0), (4,0) and (5,0) SWCNTs are separated by 0.76nm, 0.69nm and 0.61nm, respectively. As a result, the (5,0) SWCNTs holding the shortest wall-to-wall distance presumably require a longer range for the lateral interaction to vanish completely. The study of the electronic DOS as a function of energy in Figure 5 provides a guideline to increase the $T_c$ via tuning the electron concentration. Considering the best sample in zigzag family: the (4,0) SWCNTs, the increase of $T_c$ should be done by decreasing the Fermi energy by 0.27eV.

An infinitely long carbon chain is expected to be linear [27,28]. However, the situation is changed by the influence from the surrounding (5,0) SWCNTs, as shown in the relaxed structure in Figure 6. This mystery can be explained by the variation of the local atomic density on the SWCNT. Any SWCNT is made of the connection of $C_6$ rings. The central or the middle point of the $C_6$ ring holds the lowest particle density. Therefore, the interaction between the (5,0) SWCNT and the carbon nano-chain is microscopically non-uniform along the angular plane and eventually varies the kink angle and bond lengths of the inner carbon nanowire. Despite the (5,0) SWCNTs being metallic, the alternative bond lengths in the linear carbon chain may contribute to a semiconducting behavior [27], and therefore it is necessary to verify whether the composite is metallic or not. The DFT calculation confirms the metallic state of the composite. The theoretical $T_c$ of the carbon composite, is 2.2 times higher than the pure (5,0) SWCNT array of identical tube-to-tube separation. Our calculation confirms that doping elements inside the carbon nanotube should be the key to enhance their $T_c$ with help of a modification of the electron-phonon interaction. The onset theoretical $T_c$ of the mentioned SWCNTs and composite refers to the temperature to form Cooper pairs only. However, the electrical resistance does not vanish until the superconducting order parameter accomplishes a long-range phase-coherent state over the entire nanotube array [5,6]. This requires a finite Josephson interaction between adjacent nanotubes to trigger a phase-ordering transition. The difference between the onset superconducting transition temperature and the phase ordering temperature is another challenging task, which we plan to address in a future work. The scale factor approach is appropriate to the SWCNT system because its Debye temperature is much larger than the $T_c$ [29].

**Conclusion:**

We have demonstrated the theoretical limit of the maximum $T_c$ that can be reached in various ultrathin single walled carbon nanotube arrays by using a novel semi-phenomenological scale factor approach. The $T_c$ of the optimized single walled carbon nanotube array can be increased up to a maximum value of 60K and the superconductivity depends sensitively on the nanotube symmetry. The further investigation of the lateral tube-to-tube coupling provides a guideline for arranging the single walled carbon nanotubes in order to suppress the effect of phase fluctuations so that a macroscopically phase-coherent zero electrical resistance state can be reached. We have furthermore confirmed that inserting a linear carbon chain into a single walled carbon nanotubes is a promising way to further enhance their $T_c$.

**Author contribution:**

A. F. Zatsepin and E. A. Buntov and C. H. Wong planned the project. C. H. Wong designed the semi-phenomenological scale factor approach and performed the DFT calculation. A. F. Zatsepin, E. A. Buntov and C. H. Wong discussed and analyzed the data. R. E. Kasimova calculated the phonon density of states with help of C. H. Wong. All authors wrote the manuscript.

**Acknowledgments**

We acknowledge the technical support of the ab-inito software from Prof. Jiyan Dai in The Hong Kong Polytechnic University. The research was partially supported by Act 211 of the Government of the Russian Federation, contract no. 02.A03.21.0006.

**Competing financial interests**

The authors declare no competing financial interests.